\documentclass[reprint]{revtex4-1}

\usepackage{upgreek,graphicx}

\begin{document}
\title{Overcoming the speckle correlation limit to achieve a fiber wavemeter with attometer resolution}

\author{Graham D. Bruce}
\email{gdb2@st-andrews.ac.uk}
\author{Laura O'Donnell}
\author{Mingzhou Chen}
\author{Kishan Dholakia}

\affiliation{SUPA, School of Physics and Astronomy, University of St Andrews, North Haugh, St Andrews KY16 9SS, UK}
\date{February 8, 2019}
\begin{abstract}
The measurement of the wavelength of light using speckle is a promising tool for the realization of compact and precise wavemeters and spectrometers. However, the resolution of these devices is limited by strong correlations between the speckle patterns produced by closely-spaced wavelengths. Here, we show how principal component analysis of speckle images provides a route to overcome this limit. Using this, we demonstrate a compact wavemeter which measures wavelength changes of a stabilized diode laser of 5.3\,am, eight orders of magnitude below the speckle correlation limit.
\end{abstract}

\maketitle

The scattering of coherent light by a disordered medium converts the input field's spatial profile into a granular speckle pattern. Much research is devoted to minimizing or removing this effect. However, the precise speckle pattern produced is strongly dependent on the properties of both the light field and the scattering medium. As a result, speckle has been widely used as a sensor for changes in the scattering medium \cite{Yamaguchi81,Boas10,Briers13}. Alternatively, as the coherence of the light field is preserved, the speckle pattern can be used to gather information about the incoming beam if the scatterer is fixed in time. This has been harnessed in focussing, micromanipulation and imaging through turbid media \cite{Vellekoop07,Vellekoop10,Cizmar10} and in measuring the polarization \cite{Kohlgraf10}, wavelength \cite{Redding12,Redding13,Redding13b,Mazilu14,Redding14,Redding14b,Chakrabarti15,Wan15,Liew16,Metzger17,Cao17} and modal character \cite{Mazilu12,Mourka13} of the light. Precision measurement of the wavelength of light is fundamental to many fields of science, including laser spectroscopy, optical sensing and telecommunications \cite{Demtroder13}. The most common spectrometers use diffraction to spatially separate the wavelength components of light in a one-to-one spectral to spatial mapping, with the resolution therefore depending on the separation of diffracting element and detector. This one-to-one mapping is, however, not a prerequisite for spectral recovery, and tracking wavelength via speckle translates this measurement into a two dimensional problem, allowing for both high resolution and large bandwidth in an intrinsically compact design \cite{Cao17}. 

A particularly appealing method for generating laser speckle is a multi-mode optical fiber (MMF), due to its low cost, ability to transport light over long distances with minimal losses and small footprint \cite{Hecht15}. When coherent light traverses such a fiber, multiple-scattering and modal interference produce a speckle pattern. The predominant method to measure wavelength using speckle is via the transmission matrix method \cite{Cao17}. In brief, a transmission matrix $T$ is first constructed by recording the speckle patterns produced for a set of known wavelengths, where each column of $T$ contains the intensity pattern produced by a different wavelength. The MMF is stabilized so that $T$ is time-invariant. Provided that the wavelengths used to construct the transmission matrix produced uncorrelated speckle patterns \cite{Redding13b}, $T$ can subsequently be used as a reference from which one can extract the unknown wavelength that produced a measured speckle pattern. Using this method, compact spectrometers with an operating range of $\sim1\upmu$m have been realized \cite{Redding14}. 

\begin{figure}[!hb]
\centering
\includegraphics[width=0.9\linewidth,trim={6.5cm 12cm 6.5cm 12cm}]{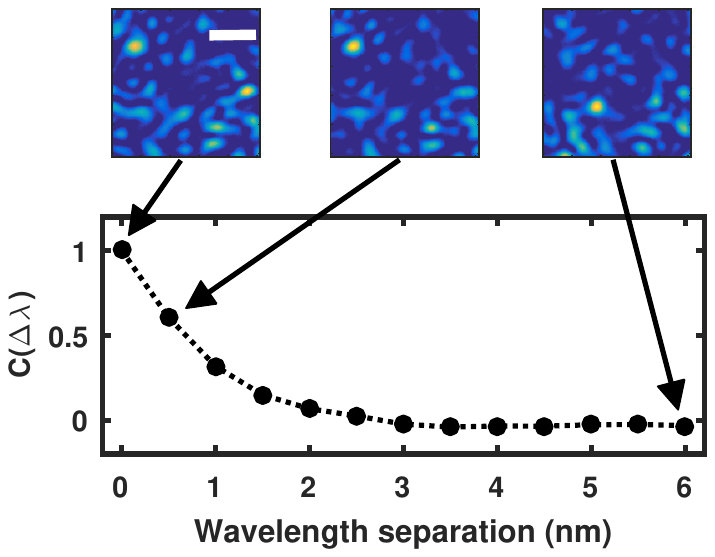}
\caption{The speckle patterns produced by an 18\,cm length of multi-mode fiber vary with wavelength. Strong correlations are present for wavelength changes within the speckle correlation limit (the HWHM of the correlation curve), which is $\sim 620$\,pm in our system. The white scale bar is common to all images and denotes a distance of 2\,mm. \label{Fig1}}
\end{figure}

The limit to the wavelength resolution, and thus the accuracy, of this method is the requirement that the speckle patterns produced by two closely-spaced wavelengths (separated by $\Delta\lambda$) are sufficiently uncorrelated. The speckle correlation function 
\begin{equation}
C\left(\Delta\lambda\right) = \left\langle \frac{\left\langle I\left(\lambda,x\right)I\left(\lambda+\Delta\lambda,x\right) \right\rangle_{\lambda} }{ \left[ \left\langle I\left(\lambda,x\right) \right\rangle_{\lambda} \left\langle I\left(\lambda+\Delta\lambda,x\right) \right\rangle_{\lambda} \right]} - 1 \right\rangle_{x},
\end{equation}
compares the intensity $I(\lambda,x)$ of images with spatial co-ordinates $x$ recorded at a range of wavelengths $\lambda$, where $\left\langle ... \right\rangle_{\lambda}$ denotes an average over wavelengths and $\left\langle ... \right\rangle_{x}$ denotes an average over position. The values of $C$ are normalized such that $C\left(0\right)=1$ \cite{Redding13}. For clarity, we illustrate this correlation function in Fig.~\ref{Fig1} using the speckle patterns produced by an 18\,cm length of MMF injected by light from an external cavity diode laser tuned between $\lambda = 775$\,nm and 785\,nm in steps of 0.5\,nm. For large wavelength separations the speckle patterns are distinct and $C \approx 0$. However, for small wavelength changes the speckle patterns are strongly correlated, i.e. $C \approx 1$. The resolution limit of the transmission matrix method is called the \emph{speckle correlation limit} and defined as the wavelength separation for which $C=0.5$, i.e. the HWHM of the correlation curve \cite{Redding13}. In the absence of strong mode coupling within the fiber, it scales inversely with the length of the optical fiber \cite{Rawson80,Freude86,Hlubina94,Redding14}. In the example shown in Fig.~\ref{Fig1}, the speckle correlation limit is $\sim620$\,pm. 

Recent work investigating speckle in other, less widely-used, optical systems has demonstrated wavelength measurements which are well-resolved below the correlation limit \cite{Mazilu14, Metzger17}, via principal component analysis (PCA). These measurements may be either of absolute wavelength or of relative wavelength changes. The reason for this improvement has thus far not been fully explained nor placed in context. Here, we show that PCA circumvents the speckle correlation limit to the wavelength resolution because it is designed to identify the optimal basis in which to measure the changes of speckle patterns. We find that this offers not only an advantage in resolution, but also in separating wavelength-dependent changes of speckle patterns from those due to environmental changes. We demonstrate for the first time that the combination of a MMF and PCA can produce a compact, \emph{attometer-resolved} wavemeter, thereby measuring wavelength changes eight orders of magnitude below the speckle correlation limit.

\begin{figure}[!t]
\centering
\includegraphics[width=1\linewidth,trim={0 .43cm 0 0}]{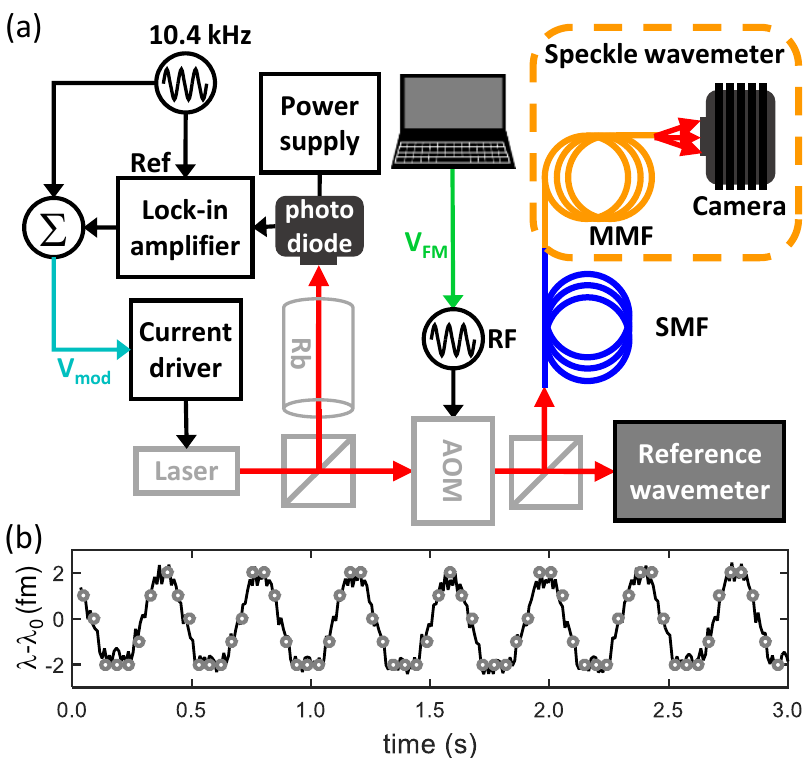}
\caption{(a) MMF speckle wavemeter and test setup. An external cavity diode laser is wavelength-stabilized using saturated absorption spectroscopy of Rb and active feedback to produce a control voltage $V_{mod}$ applied to the laser current driver. The wavelength of the light is subsequently tuned using an acousto-optic modulator (AOM) as determined by control voltage $V_{FM}$. The light is delivered to both the reference and speckle wavemeters via single-mode optical fiber (SMF). (b) Speckle (black) and reference (gray circles) wavemeter measurements for AOM modulation of 2\,fm. \label{Fig2}}
\end{figure}

We investigate the attainable resolution of speckle-based wavelength measurement using a speckle wavemeter comprising an 18\,cm-long step-index MMF and a fast CMOS camera (Fig.~\ref{Fig2}a). The MMF core diameter is 105\,$\upmu$m with NA = 0.22 (ThorLabs FG105LCA). After exiting the MMF, the light propagates for 5\,cm and is captured by the camera (Mikrotron EoSens 4CXP). Laser light for testing the speckle wavemeter is generated by an external cavity diode laser (Toptica DL-100, LD-0785-P220), which is stabilized to the $^{87}$Rb D$_2$ line ($F=2 \rightarrow F^{\prime}=2\times3$ crossover) using saturated absorption spectroscopy and current modulation. To vary the wavelength of the light, we employ an acousto-optic modulator (AOM) (Crystal Technologies 3110-120) in a cat-eye double-pass configuration, with a modulation range of 20\,fm. The wavelength tuning is set by Labview and passed to an RF function generator as control voltage $V_{FM}$. In this letter, all wavelength measurements are quoted relative to $\lambda_{0}=780.212840$\,nm. The light is coupled into an angle-cleaved single-mode fiber (SMF) (ThorLabs P5-780PM-FC-10) to eliminate spatial variations of the beam. This is connected to the MMF via a standard mating sleeve (Thorlabs ADAFC1), delivering 300\,$\upmu$W of power to the camera.

We use PCA on the images recorded on the camera to extract the wavelength, as previously detailed in Ref. \cite{Metzger17}. Briefly, a training set of independently-normalized speckle patterns at a range of known wavelengths is obtained. The principal components (PCs) of the data are given by the eigenbasis of the covariance matrix of the training set. There exists a linear relationship between the value of the first principal component PC1 (i.e. the eigenvector with largest eigenvalue) and wavelength, with the proportionality constant determined by the training set. For a speckle pattern produced by an unknown wavelength, the wavelength is extracted by projecting the speckle pattern into the PC-space of the training set.

For wavelength changes above a femtometer, we benchmarked the speckle wavemeter against a High-Finesse WS7 wavemeter (Fig.\,\ref{Fig2}b). A sinusoidal wavelength modulation with 2\,fm amplitude was applied at 2.5\,Hz using the AOM. We recorded images of $1024 \times 1024$ pixels at 100\,fps and 2\,$\upmu$s exposure. An estimation of the relative accuracy of the speckle wavemeter and the commercial wavemeter is provided by the standard deviation of the residuals between these two measurements, which was 0.28\,fm over the 3\,s measurement.

\begin{figure}[!t]
\centering
\includegraphics[width=0.6\linewidth,trim={4cm 6cm 4cm 6cm}]{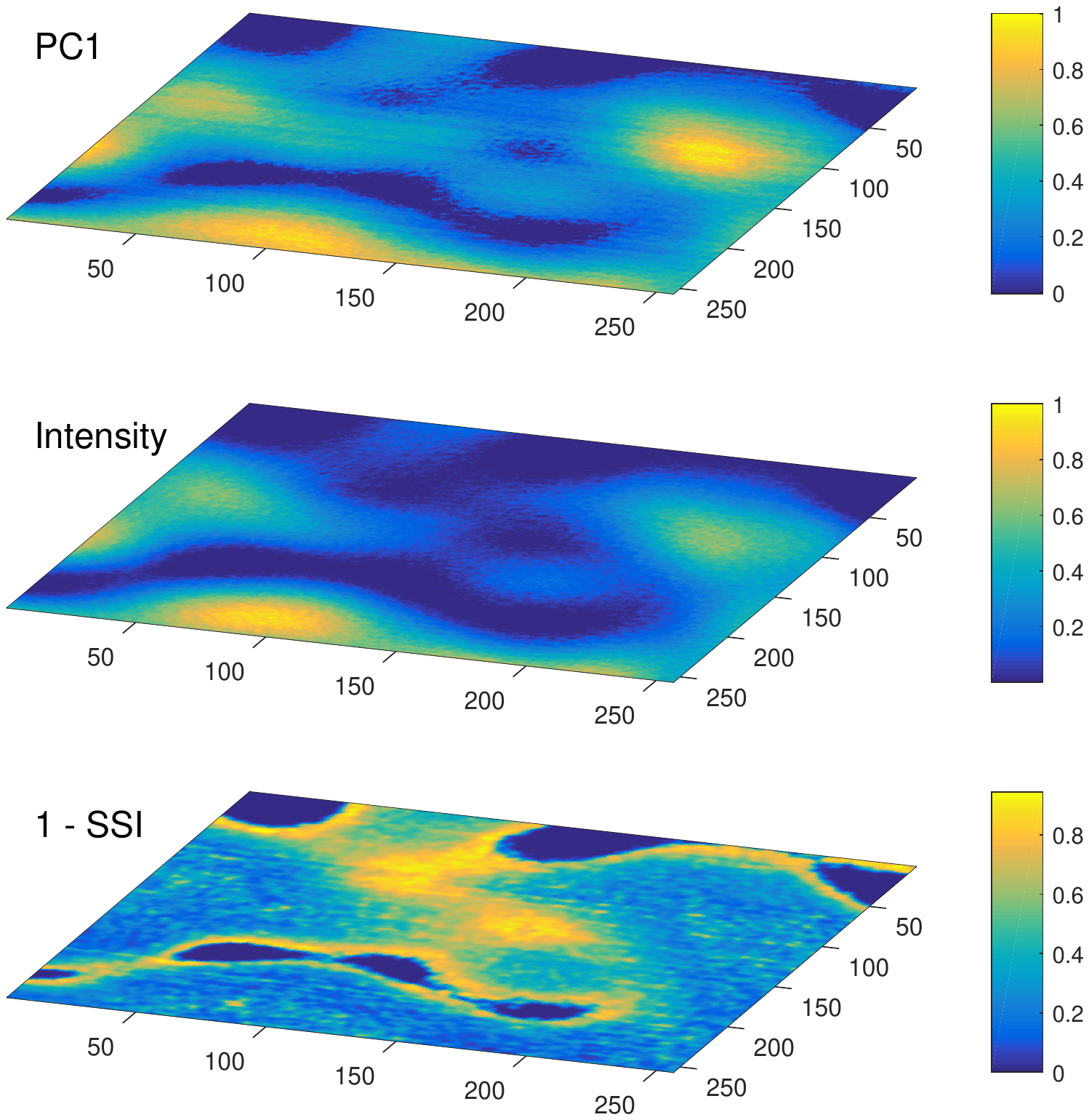}
\caption{The distribution of the pixel-weightings in PC1 (top) resembles a typical speckle pattern from the training set (middle). However, mapping the dissimilarity between these (bottom) shows that PCA gives increased weight to low-intensity pixels near the edges of the speckles. \label{Fig3}}
\end{figure}

Compared to the other scattering media to which PCA has been applied, the MMF has a relatively low numerical aperture and supports relatively few transverse optical modes. This gives rise to a coarse-grained speckle pattern which is ideally suited to illustrate the underlying mechanisms allowing measurements below the speckle correlation limit, as illustrated in Fig.~\ref{Fig3}. We find that, after the training procedure to identify the principal components, PC1 weighs the contributions of all pixels in the speckle image, but with added weight applied to low-intensity pixels at the edges of the speckle grains. This can be seen when comparing the pixel weightings of PC1 in image space to a typical speckle pattern from the training set. To highlight this finding, Fig.~\ref{Fig3} shows the structural dissimilarity , i.e. 1~-~SSI, where SSI is the structural similarity index \cite{Zhou04}. The dissimilarity takes large values at the edges of the grains. This is because as the wavelength changes, the speckles in the pattern move and change in size. Such intensity variations at the edges of patterns are well known to be disproportionately sensitive to small changes, and have been well used in, for example, extraction of sound from silent video \cite{Davis14}.

\begin{figure}[!t]
\centering
\includegraphics[width=0.85\linewidth,trim={6.5cm 9.8cm 6.5cm 9.8cm}]{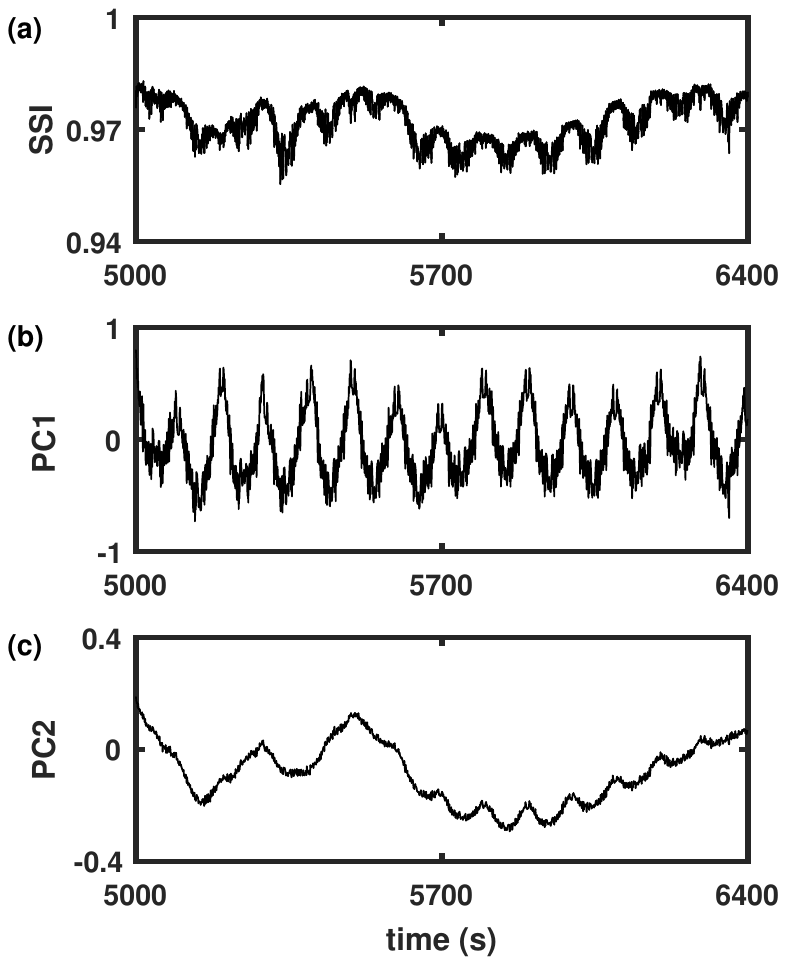}
\caption{(a) Image-based recovery of the wavelength based on parameters like structural similarity index are susceptible to corruption due to coupling between environmental- and wavelength-dependent variations of the speckle. (b-c) However, principal component analysis, by identifying the optimal basis to recover the wavelength, decouples environmental factors from the wavelength-dependent variations in the speckle. \label{Fig4}}
\end{figure}

By identifying the optimal basis for wavelength measurements, PCA can be used to improve the stability of speckle-based wavelength measurement against environmental fluctuations which would otherwise disrupt the wavelength recovery. Using a short MMF gives low sensitivity to environmental factors including temperature and vibration \cite{Redding14}, and over short timescales no change in speckle patterns can be seen for a fixed wavelength. However, when measuring over longer timescales, the speckle patterns do show sensitivity to the environment. In Fig.~\ref{Fig4}a, we show the SSI of speckle images over 6400\,s, recorded at 10\,fps and 1\,$\upmu$s exposure. Throughout the measurement, the wavelength was modulated by the AOM at 10\,mHz with an amplitude of 1\,fm. Tracking the changes in speckle through an image-based metric such as the SSI does capture the 10\,mHz modulation, but with a non-constant amplitude of oscillation and a time-dependent mean value. The amplitude of the SSI oscillation at 10\,mHz varies by one order of magnitude, from a minimum of 0.08\% to a maximum of 0.6\%, while the mean value of the SSI also varies by 5.3\%. This implies a wavelength drift at least 8.8 times larger than the control modulation. In contrast, evaluating the same speckle images with PCA shows a separation of the wavelength modulation and slower drifts in PC1 and PC2 respectively. The amplitude of the oscillations in PC1 (Fig.~\ref{Fig4}b) varies by 16\%, with a mean wavelength drift of a factor 1.1 times the modulation, consistent with observations using the reference wavemeter. The remaining changes in the speckle patterns are instead captured by higher order PCs. We highlight PC2 in Fig.~\ref{Fig4}c.

\begin{figure}[!t]
\centering
\includegraphics[width=0.85\linewidth,trim={0 .5cm 0 0cm}]{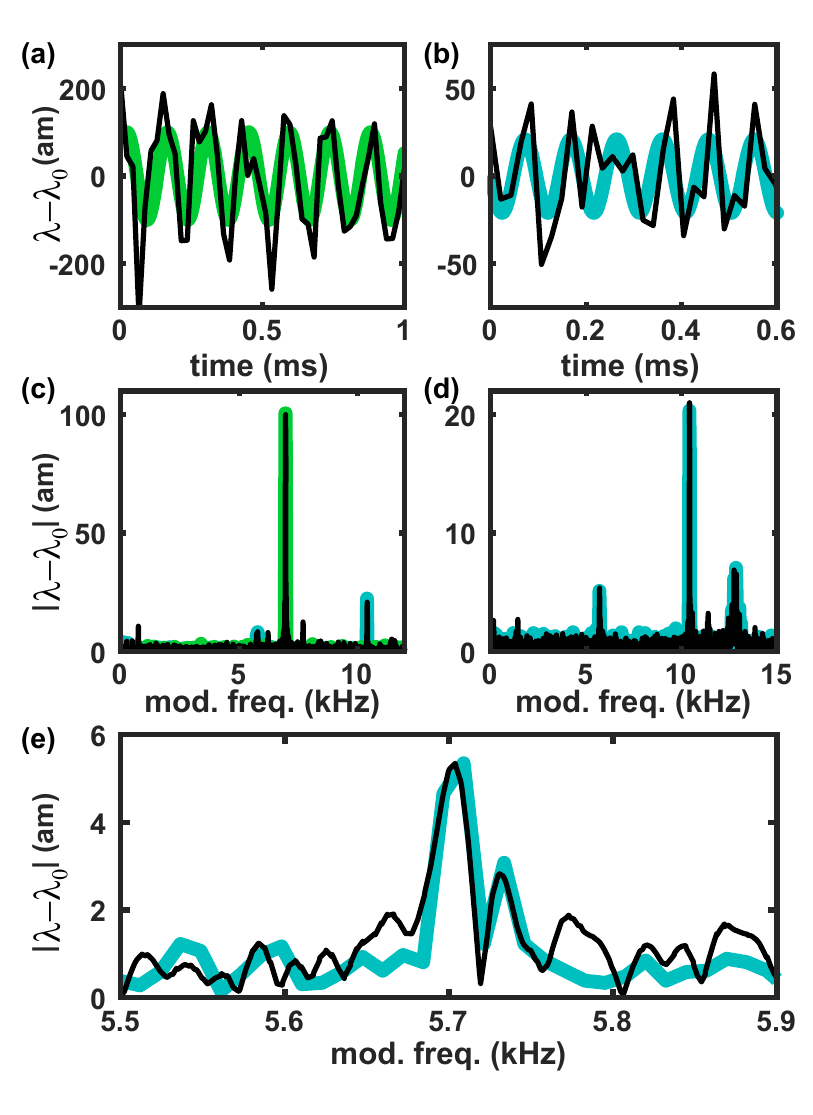}
\caption{Speckle wavemeter measurement using principal component analysis (black) for (a) AOM modulation of 0.1\,fm and (b) no modulation. (c, d) Amplitude spectra of (a, b). (e) The smallest measured signal component with amplitude of 5.3\,am. Green (blue) shows $V_{FM}$ $\left(V_{mod}\right)$. In (b), blue is a guide-to-the-eye showing the component from the 10.4\,kHz oscillator only.          \label{Fig5}}
\end{figure}
To investigate the resolution of the device, we subsequently applied wavelength modulations below femtometer-level to the light using the AOM. The calibrated modulation signal from Labview $V_{FM}$ (highlighted in green in Fig.\,\ref{Fig2}a) was $100 \pm 4$\,am at $6.97 \pm 0.03$\,kHz. We acquired 5,000 frames of $8\times 1024$ pixels, with $3\,\upmu$s exposure at 48,000\,fps. The response of the speckle wavemeter, shown in Fig.\,\ref{Fig5}a clearly resolves this wavelength modulation, but with notable deviations from the sinusoidal modulation signal. Each data point is a measurement of the instantaneous wavelength, extracted from a single image. However, computing the amplitude spectrum of the complete image set allows us to illustrate why discrepancies occur between the control modulation and the response of the speckle wavemeter, and to test the accuracy and precision of the wavemeter. The amplitude spectrum (Fig.\,\ref{Fig5}c) of the speckle wavemeter signal shows a dominant peak at $6.96 \pm 0.01$\,kHz (which is caused by the control voltage) along with additional, smaller, modulation components. Using the dominant peak as a benchmark, we determined that the second peak in the spectrum is caused by an additional sinusoidal wavelength modulation at $10.42 \pm 0.01$\,kHz with an amplitude of $20.7 \pm 0.7$\,am. This modulation was caused by the external oscillator signal (dither) used to wavelength-stabilize the laser. We verified our wavelength measurements against the modulation voltage $V_{mod}$ (shown in blue in Fig.\,\ref{Fig2}a) applied to the laser diode current driver (Thorlabs LDC-202B). The wavelength dependence on $V_{mod}$ was $53\pm 4$\,am\,mV$^{-1}$, and the predicted wavelength modulation due to the dither was therefore $22 \pm 3$\,am. For the benchmarking voltage measurements, the quoted uncertainties are the combination of 5\% calibration error and the standard error of five repetitions. The relative accuracy of the speckle wavemeter is within this errorbar, and the precision of the speckle wavemeter is on attometer scale.

To further demonstrate attometer-resolved measurements with the speckle wavemeter, we recorded a time-series with no AOM modulation (Fig.~\ref{Fig5}b). The amplitude spectrum (Fig.~\ref{Fig5}d) of the recorded signal shows wavelength modulations occurring at three distinct frequencies. The largest of these was the dither, which was used to benchmark the additional modulation components. A single tone at a modulation frequency of $5.70 \pm 0.02$\,kHz and a 0.4\,kHz-broad, multipeak feature centered at 12.8\,kHz were caused respectively by a noise component in the lock-in amplifier and the switch-mode power supply of the photodetector shown in Fig.~\ref{Fig2}. The lock-in amplifier circuitry produced a component to $V_{mod}$ at $5.7 \pm 0.3$\,kHz with $V_{pk}=96 \pm 3\,\upmu$V, giving a wavelength modulation of $5.1\pm 0.5$\,am. This matches the $5.3 \pm 0.4$\,am measurement obtained with the speckle wavemeter (Fig.~\ref{Fig5}e). This measurement is well-resolved, with a peak larger than $10\sigma$, where $\sigma$ is the standard deviation of the noise floor. The voltage noise from the switch-mode supply was added to the photodetector signal, and thus also to $V_{mod}$. The most prominent component of this feature had an amplitude of $V_{pk}=133 \pm 4\,\upmu$V, i.e. a $7.1\pm 0.7$\,am wavelength modulation. The measurement from the speckle wavemeter was in agreement, at $6.5 \pm 0.7$\,am. 

In conclusion, we have demonstrated wavelength measurements which are well-resolved despite being eight orders of magnitude below the speckle correlation limit, through the use of PCA. This approach can be used in tandem with existing transmission matrix methods to maintain broadband performance. At present the resolution is limited by the white noise floor of our experiment, and intriguingly it is invariant for fiber lengths up to 50\,m. A detailed analysis of the limits of PCA will be the subject of future work, but promisingly the length-invariance suggests that the approach may suitable for on-chip integrated photonics. The observation that environmental and wavelength changes to the speckle patterns are separated between principal components raises the possibility that, by implementation of an appropriate training process, the speckle wavemeter could be used as a multi-parameter sensor. Additionally, we will test whether the approach can be extended from measurements of narrow-linewidth, monochromatic laser light into a spectrometer capable of measuring multiple wavelengths or spectra.

\begin{acknowledgements}
We acknowledge funding from Leverhulme Trust (RPG-2017-197) and EPSRC (EP/R019541/1, EP/R004854/1) and thank Donatella Cassettari, Philip Ireland and Paloma Rodr\'iguez-Sevilla for technical assistance and useful discussions. Research data supporting this publication can be accessed at \url{https://doi.org/10.17630/c567afee-de2a-4c8c-9413-0f0f1bb8df64}.
\end{acknowledgements}

\bibliography{MMF_Speckle_Wavemeter_bib}

\end{document}